\documentclass[%
 reprint,
 amsmath,amssymb,
 aps,
 pra,
]{revtex4-2}
\usepackage{graphicx}% Include figure files
\usepackage{dcolumn}% Align table columns on decimal point
\usepackage{bm}% bold math

\usepackage{hyperref}% add hypertext capabilities
\begin{document}

%\preprint{APS/123-QED}

\title{Multimode parity-time and loss-compensation symmetries in coupled waveguides with loss and gain}

\author{Anton~V.~Hlushchenko}
\affiliation{State Key Laboratory of Integrated Optoelectronics, College of Electronic Science and Engineering, International Center of Future Science, Jilin University, 2699 Qianjin Street, Changchun, 130012, China}
\affiliation{National Science Center `Kharkiv Institute of Physics and Technology' of National Academy of Sciences of Ukraine, 1, Akademicheskaya Street, Kharkiv 61108, Ukraine}
\author{Vitalii~I.~Shcherbinin}
\affiliation{National Science Center `Kharkiv Institute of Physics and Technology' of National Academy of Sciences of Ukraine, 1, Akademicheskaya Street, Kharkiv 61108, Ukraine}
\author{Denis~V.~Novitsky}
\affiliation{B. I. Stepanov Institute of Physics, National Academy of Sciences of Belarus, 68 Nezavisimosti Avenue, Minsk 220072, Belarus}
\author{Vladimir~R.~Tuz}
\email{tvr@jlu.edu.cn}
\affiliation{State Key Laboratory of Integrated Optoelectronics, College of Electronic Science and Engineering, International Center of Future Science, Jilin University, 2699 Qianjin Street, Changchun, 130012, China}

\date{\today}

\begin{abstract}
Loss compensation via inserting gain is of fundamental importance in different branches of photonics, nanoplasmonics, and metamaterial science. This effect has found an impressive implementation in the parity-time symmetric ($\mathcal{PT}$-symmetric) structures possessing balanced distribution of loss and gain. In this work, we generalize this phenomenon to the asymmetric systems demonstrating loss compensation in the coupled multi-mode loss-gain dielectric waveguides of different radii. We show that similar to the $\mathcal{PT}$-symmetric coupled single-mode waveguides of identical radii, the asymmetric systems support the exceptional points called here the loss compensation (LC) thresholds where the frequency spectrum undergoes a transition from complex to real values. Moreover, the LC-symmetry thresholds can be obtained for dissimilar modes excited in the waveguides providing an additional degree of freedom to control the system response. In particular, changing loss and gain of asymmetric coupled waveguides, we observe loss compensation for TM and TE modes as well as for the hybrid HE and EH modes.
\end{abstract}

\maketitle

%\tableofcontents

\section{Introduction}

The intriguing properties of loss-gain distributions are the subject of the fast-growing field of non-Hermitian photonics. The most notable and studied class of non-Hermitian systems is the $\mathcal{PT}$-symmetric systems which possess the perfectly balanced spatial distribution of gain and loss \cite{Zyablovsky_UFN_2014, Feng_NatPhot_2017, El-Ganainy_NatPhys_2018}. $\mathcal{PT}$ symmetry guarantees the reality of non-Hermitian Hamiltonian spectra \cite{Bender_PRL_1998} that have been readily implemented and observed in the optical domain \cite{Makris_PRL_2008, Ruter_NatPhys_2010}. Subsequently, $\mathcal{PT}$ symmetry was transferred to electronic circuits \cite{Schindler_PRA_2011}, acoustics \cite{Zhu_PRX_2014}, and time-varying (Floquet) systems \cite{Chitsazi_PRL_2017}. 

The balance between loss and gain helps to solve the problem of loss compensation which is of paramount importance for the efficiency of plasmonic and metamaterial-based devices \cite{Hess_NatMat_2012, Krasnok_IEEE_2020, Novitsky_PRB_2017}. The loss compensation effect proved to be connected to another remarkable feature of non-Hermitian systems -- the possibility of specific degeneracies called exceptional points (EPs) which were observed in photonics \cite{Ozdemir_NatMat_2019, Miri_Sci_2019} as well as in acoustics \cite{Kun_PRX_2016}. In contrast to the degeneracies of Hermitian systems (the so-called diabolic points) with coalescing eigenvalues, the EPs imply coalescence of both eigenvalues and eigenfunctions. In the $\mathcal{PT}$-symmetric systems, the EPs mark the points of phase transitions between the $\mathcal{PT}$-symmetric phase and the phase with spontaneously broken $\mathcal{PT}$ symmetry. It is important to emphasize, however, that the EPs are the general phenomenon observed also in purely passive systems (such as whispering-gallery-mode microresonators \cite{Jiang_PRA_2020}, ring cavities \cite{Wang_PRA_2020}, and anisotropic waveguides \cite{Gomis-Bresco_PRR_2019}) and even in the systems with radiative loss only \cite{Abdrabou_JOSAB_2019}.

The EPs have become a workhorse of many recent achievements such as enhanced perturbation sensing \cite{Chen_Nat_2017, Hodaei_Nat_2017}, novel lasing schemes \cite{Feng_Sci_2014, Hodaei_Sci_2014}, enhanced Sagnac effect for laser gyroscopes \cite{Hokmabadi_Nat_2019, Lai_Nat_2019}, simultaneous coherent perfect absorption and amplification \cite{Longhi_PRA_2010, Wong_NatPhot_2016}, asymmetric transmission \cite{Makris_PRL_2008, Novitsky_PRB_2018}, etc. An interesting feature of EPs is their topological nature which can be revealed with their dynamical encircling in parameter space and can be used for mode switching \cite{Doppler_Nat_2016}, mode transfer \cite{Liu_PRL_2020}, and polarization conversion \cite{Hassan_PRL_2017}. Another exciting direction is the observation and utilization of the higher-order EPs, where more than two eigenmodes coalesce. Such EPs can be realized either in the systems containing three or more resonant elements \cite{Hodaei_Nat_2017, Wang_NatComm_2019, Zhong_PRL_2020} or by hybridizing several usual (second-order) EPs \cite{Ryu_PhotRes_2019, Zhang_CommPhys_2019}. It should be noted that the most considerations of non-Hermitian effects are limited to the single-mode case with the use of the coupled-mode theory (CMT) \cite{Huang_1994}. The multi-mode platforms provide much richer opportunities in controlling optical response and dynamics as evidenced by the examples of dispersion engineering and mode conversion in the waveguide and microresonator systems \cite{Dai_OE_2012, Zhang_SciRep_2015, Kim_NatComm_2017}.

For the symmetric system of coupled dielectric waveguides, the full loss compensation is possible only for the perfect balance between gain and loss corresponding to the non-violated $\mathcal{PT}$-symmetry \cite{Ruter_NatPhys_2010}. This poses rather tough and difficult-to-achieve conditions for experimental realization of the related effects \cite{Ozdemir_NatMat_2019}. On the contrary, for the asymmetric system, the full loss compensation can be reached even for the unbalanced gain and loss. By analogy with the $\mathcal{PT}$-symmetry, this generalized situation can be called the loss-compensation (LC) symmetry \cite{Klimov_LaserPhysLett_2018, Hlushchenko_LaserPhysLett_2020}. In this paper, we propose further generalization demonstrating LC symmetry for the coupled dissimilar waveguides under excitation of the modes with the same or different azimuthal indices. In dielectric waveguides, all modes except TE and TM are hybrid, i.e., they have axial components of both electric and magnetic fields. Therefore, to analyze the system, we apply the multi-mode approach \cite{White_JOSAB_2002, Hlushchenko_LaserPhysLett_2020} which allows us to obtain exact solutions of the eigenvalue problem for all possible classes of modes supported by the asymmetric guiding structure. We show that such an asymmetric structure supports specific EPs called the LC-symmetry thresholds and reached for different loss-gain ratios depending on the modes used. Thus, the mode composition of the non-Hermitian system is an additional degree of freedom useful for tuning the parameters necessary to obtain the constant-intensity modes \cite{Kominis_SciRep_2016, Kominis_PRA_2017} and other applications.

%In cylindrical dielectric waveguides, all modes except TE and TM are hybrid, i.e., they have axial components of both electric and magnetic fields.

\section{Multimode analytical approach}

\begin{figure}
\includegraphics[width=\linewidth]{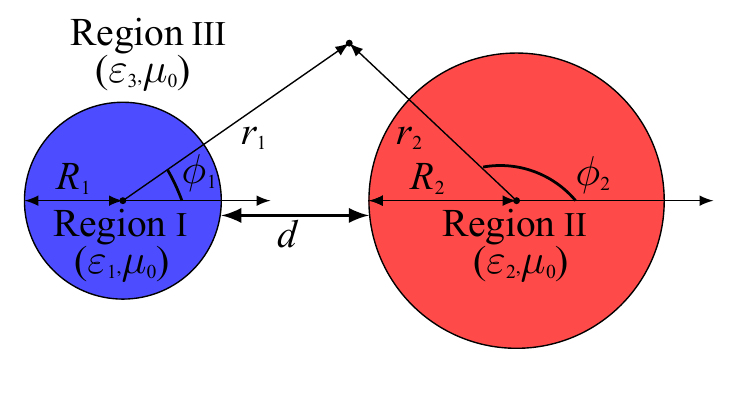}
\caption{\label{fig_cylinders} Geometry of unequally sized dielectric waveguides immersed in an infinite medium.}
\end{figure}

For our analysis, we use the multi-mode analytical approach \cite{White_JOSAB_2002}, which was previously successfully applied to coupled systems with loss and gain \cite{Hlushchenko_LaserPhysLett_2020}. Following Ref. \cite{Hlushchenko_LaserPhysLett_2020}, we consider a pair of coupled dielectric cylinders of the radii $R_1$ and $R_2$ placed in an ambient medium. Figure \ref{fig_cylinders} shows the cross-sections of the dielectric circular waveguides with the permittivities $\varepsilon_1$ and $\varepsilon_2$, respectively (Regions I and II). The polar coordinate systems associated with the waveguides are denoted as ($r_1, \phi_1$) and ($r_2, \phi_2$). The waveguides are placed in the infinite uniform medium (Region III) with the real permittivity $\varepsilon_3 \in\Re$. The permeability $\mu=\mu_0$ is assumed to be the same in all regions. Without loss of generality, we suppose that the first waveguide contains gain material and the second one contains lossy medium (Im$(\varepsilon_1)<0$ and Im$(\varepsilon_2)>0$), whereas Re$(\varepsilon_1)>\varepsilon_3$ and Re$(\varepsilon_2)>\varepsilon_3$. 

In order to solve the eigenvalue problem for this system, we write the electromagnetic fields in every region using the corresponding polar coordinates as shown in Fig. \ref{fig_cylinders}. For example, the $E_z$ and $H_z$ components of electromagnetic fields in the Region~I are written in terms of the local coordinates $(r_1, \phi_1)$ as
\begin{equation}\label{field_1}
\begin{split}
E_z^1&=\sum_{n=-N}^{N}{A_n^1J_n(k_{p,1}r_1)e^{in\phi_1}}, \\ H_z^1&=\sum_{n=-N}^{N}{B_n^1J_n(k_{p,1}r_1)e^{in\phi_1}},
\end{split}
\end{equation}
whereas in the Region~II the local coordinates $(r_2, \phi_2)$ are used as
\begin{equation}\label{field_2}
\begin{split}
E_z^2&=\sum_{n=-N}^{N}{A_n^2J_n(k_{p,2}r_2)e^{in\phi_2}}, \\ H_z^2&=\sum_{n=-N}^{N}{B_n^2J_n(k_{p,2}r_2)e^{in\phi_2}},
\end{split}
\end{equation}
and, finally, in the Region~III we have the contributions from both local coordinate systems,
\begin{equation}\label{field_3}
\begin{split}
E_z^3=&\sum_{n=-N}^{N}{C_n^{1}H_n^{(1,2)}(k_{p,3}r_1)e^{in\phi_1}} \\
    & +\sum_{n=-N}^{N}{C_n^2H^{(1,2)}_n(k_{p,3}r_2)e^{in\phi_2}},\\
H_z^3=&\sum_{n=-N}^{N}{D_n^1H^{(1,2)}_n(k_{p,3}r_1)e^{in\phi_1}} \\
& +\sum_{n=-N}^{N}{D_n^2H^{(1,2)}_n(k_{p,3}r_2)e^{in\phi_2}}.
\end{split}
\end{equation}
Here $\{A_n^1, A_n^2, B_n^1, B_n^2, C_n^1, C_n^2, D_n^1, D_n^2\}$ are the unknown amplitudes of azimuthal harmonics, $k_{p,i}^2=k^2_i-k_z^2$, $k_i=k_0\varepsilon_{r_i}$  $(i=1,2,3)$, $\varepsilon_{r_i}=\varepsilon_i/\varepsilon_0$ is the relative permittivity, $k_0^2=\omega^2\varepsilon_0\mu_0$, $J_n(\cdot)$ is the Bessel function, $H_n^{(1)}(\cdot)$ and $H_n^{(2)}(\cdot)$ are the Hankel functions of the first and second kind, respectively, and the field factor of the form $\exp{[-i(\omega t - k_z z)]}$ is assumed and omitted. The choice of the Hankel function is governed by the boundary conditions for the fields in the Region III, as follows: $E_z^3 \rightarrow 0$ and $H_z^3 \rightarrow 0$ for $r_1, r_2 \rightarrow \infty$.

Since the derivation of other field components is cumbersome, it is relegated to the Appendix. We only note that the Maxwell equations and Eqs. (\ref{field_1})-(\ref{field_3}) allow us to obtain the expressions for the electromagnetic field components $E_{\phi}$ and $H_{\phi}$ in every Region. The unknown coefficients $\{A_n^1, A_n^2, B_n^1, B_n^2, C_n^1, C_n^2, D_n^1, D_n^2\}$ and $k_z$ can be obtained from the dispersion relation derived using the continuity conditions for the tangential fields at the boundary surfaces $r_1 = R_1$ and $r_2 = R_2$ (see Appendix).

\section{TM and TE modes}

\subsection{Single waveguide modes}

\begin{figure}
\includegraphics[width=\linewidth]{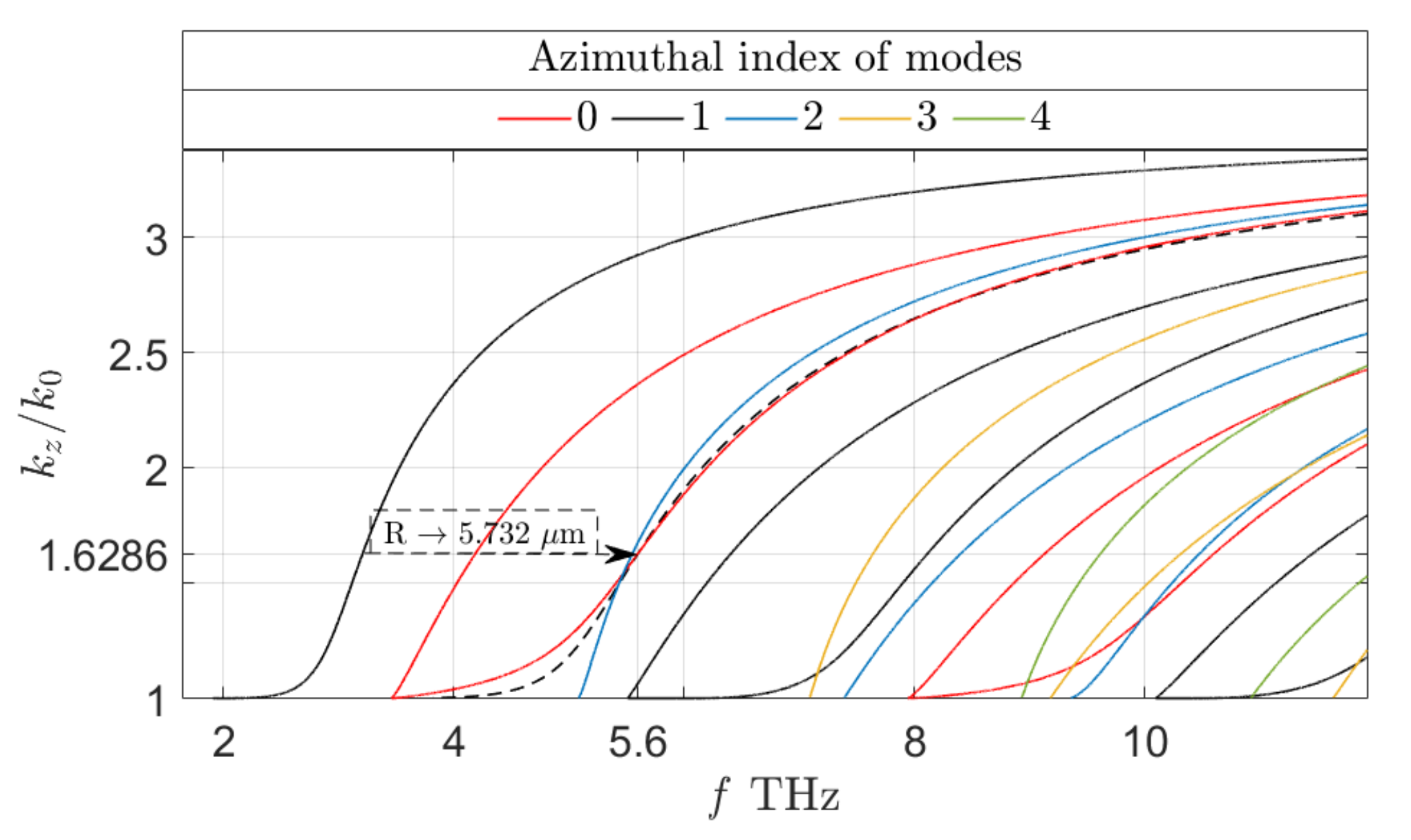}
\caption{\label{fig_full_modes} Lowest-order modes of the circular dielectric waveguide with $\epsilon_r=12$ and $R=10$~$\mu$m}
\end{figure}

Our aim is to demonstrate the possibility of full loss compensation for the coupled dielectric waveguides with gain and loss supporting different modes with an arbitrary value of azimuthal index. The dispersion relations of coupled waveguides will tend to the dispersion relations of independent waveguides with increasing distance between them. Therefore, we are interested in the crossing points of the dispersion curves of two independent cylinders. The dispersion curves for the lowest-order modes of a single waveguide are shown in Fig. \ref{fig_full_modes}. Changing the radius or permittivity of another cylinder, one can observe the shift of its dispersion curves with respect to the curves of the first one. Thus, for any pair of modes, we can find their crossing points at a required frequency $f$ and longitudinal wavenumber $k_z/k_0$. We will show further that these crossing points are convenient for the realization of full loss compensation at the certain values of gain-loss parameter and distance between the cylinders.

Let us consider, for the moment, the family of transverse electric TE$_{0m}$ and transverse magnetic TM$_{0m}$ modes of the circular cylindrical waveguide. As representatives of these modes, we choose the set \{TM$_{01}$, TM$_{02}$, TE$_{01}$, TE$_{02}$\}. For example, for the TM$_{01}$ (TE$_{01}$) mode, we can choose an arbitrary solution of the dispersion relation ($f$; $k_z/k_0$). Then, we increase the cylinder radius to obtain same solution for the TM$_{02}$ (TE$_{02}$) mode. The initial and increased radii of the waveguide can serve as a first approximation to find the conditions of loss compensation for the coupled cylinders with gain and loss. For instance, according to Fig. \ref{fig_full_modes}, the values $f=5.6$ THz and $k_z/k_0=1.6286$ correspond to the TM$_{01}$ (TE$_{01}$) mode of the waveguide with $\epsilon_{r1}=12$ and $R_1=10$ $\mu$m and to the TM$_{02}$ (TE$_{01}$) mode of the waveguide with $\epsilon_{r2}=12$ and $R_2=18.93$ $\mu$m ($R_2=16.305$ $\mu$m).

\subsection{Symmetric coupled waveguides: $\mathcal{PT}$ symmetry}

\begin{figure}
\includegraphics[width=\linewidth]{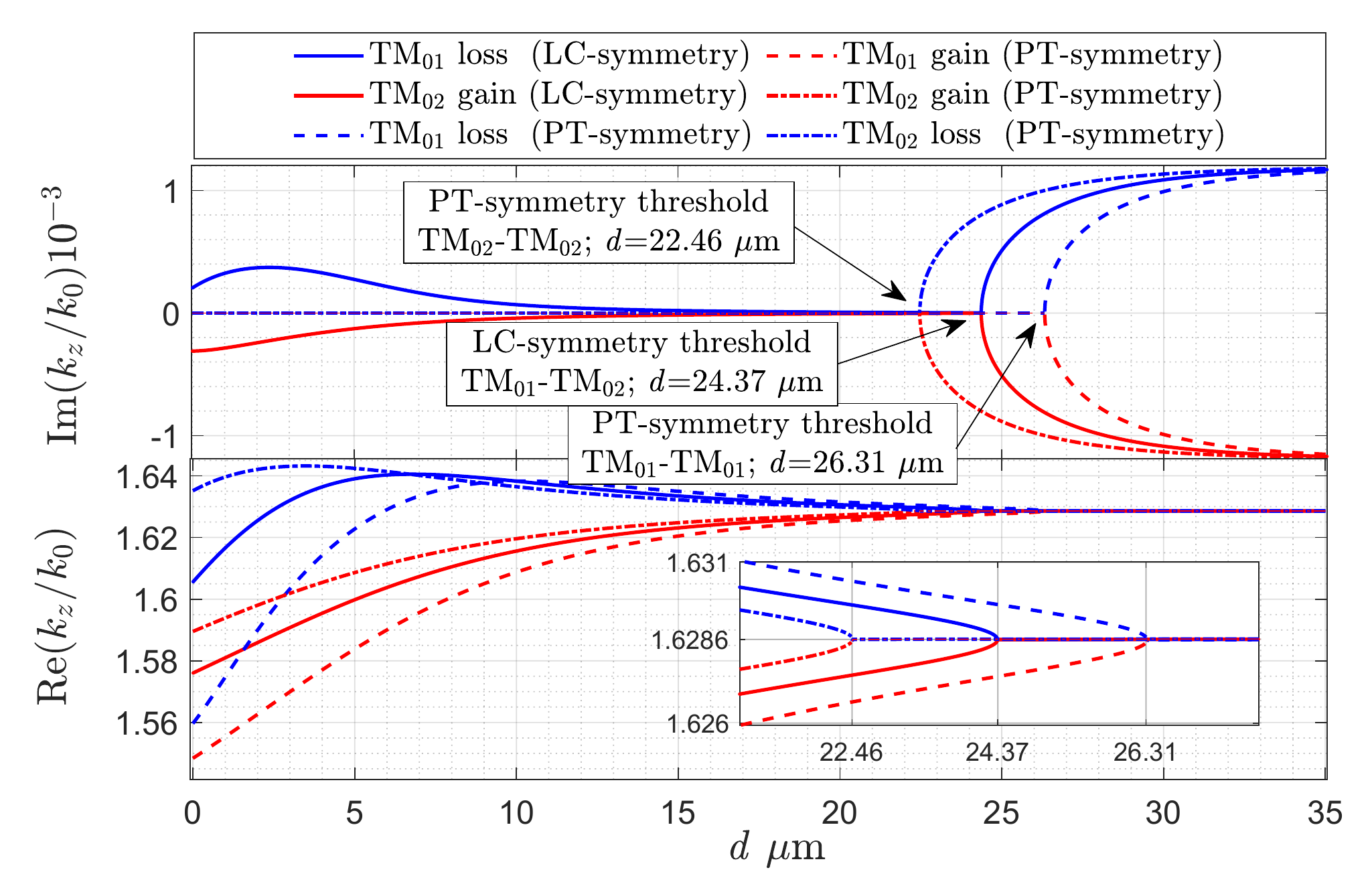}
\caption{\label{fig_coupling_tm} The real and imaginary parts of the eigenvalues $k_z/k_0$ for the TM modes as a function of the distance $d$ between the coupled waveguides. LC- and $\mathcal{PT}$-symmetry thresholds of TM-type modes are shown with arrows.}    
\includegraphics[width=\linewidth]{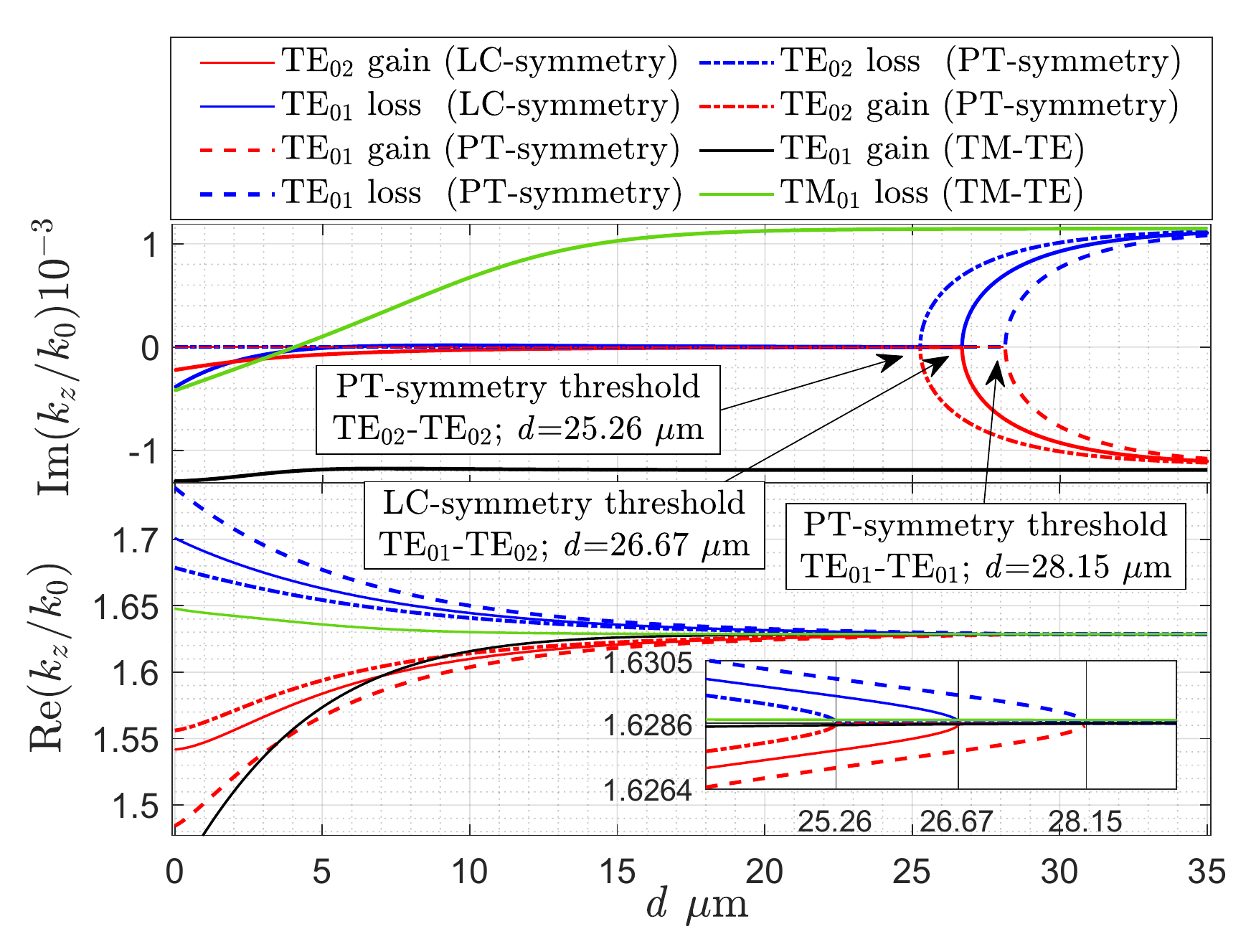}
\caption{\label{fig_coupling_te} The real and imaginary parts of the eigenvalues $k_z/k_0$ for the TE modes as a function of the distance $d$ between the coupled waveguides. LC- and $\mathcal{PT}$-symmetry thresholds of TM-type modes are shown with arrows.}
\end{figure}

We start with the case of symmetric system consisting of the coupled waveguides with the same radii and permittivities, i.e., $R_1=R_2$ and $\epsilon_{r1}=\epsilon_{r2}$. In the case of balanced loss and gain, when the loss tangents have the same absolute value, $\tan{\delta_1}=-\tan{\delta_2}$, this system is a $\mathcal{PT}$-symmetric one. $\mathcal{PT}$-symmetric systems can exist in two states: a $\mathcal{PT}$-symmetric state with the loss exactly compensated by gain, and the broken-$\mathcal{PT}$-symmetry state with the violated compensation. In our case, the transition between these states can be realized by changing the distance between the cylinders. The point of transition where the $\mathcal{PT}$ symmetry gets broken is the EP, where the modes of the system are degenerate. For example, in Fig. \ref{fig_coupling_tm}, the dashed and dash-dotted lines correspond to the TM$_{01}$ and TM$_{02}$ modes of the coupled waveguides with balanced gain and loss (the parameters are $R_1=R_2=10$ $\mu$m, $\epsilon_{r1}=\epsilon_{r2}=12$, $\tan{\delta_1}=-\tan{\delta_2}=5\cdot10^{-4}$ for TM$_{01}$ and $R_1=R_2=18.93$ $\mu$m, $\epsilon_{r1}=\epsilon_{r2}=12$, $\tan{\delta_1}=-\tan{\delta_2}=4.05\cdot10^{-4}$ for TM$_{02}$), whereas in Fig. \ref{fig_coupling_te}, the dashed and dash-dotted lines correspond to the TE$_{01}$ and TE$_{02}$ modes of the system with another set of parameters ($R_1=R_2=7.48$ $\mu$m, $\epsilon_{r1}=\epsilon_{r2}=12$, $\tan{\delta_1}=-\tan{\delta_2}=5\cdot10^{-4}$ for TE$_{01}$ and $R_1=R_2=16.305$ $\mu$m, $\epsilon_{r1}=\epsilon_{r2}=12$, $\tan{\delta_1}=-\tan{\delta_2}=4.09\cdot10^{-4}$ for TE$_{02}$, respectively). The EPs are denoted as the $\mathcal{PT}$-symmetry thresholds, since below these points (for distances smaller than the threshold one) the eigenvalues $k_z/k_0$ become purely real and the system falls into the $\mathcal{PT}$-symmetric state. This state is preserved to the very connection of the cylinders at $d=0$. It is important to note that due to the symmetry of the system, the full loss compensation via the $\mathcal{PT}$-symmetric state formation is observed only for the identical modes of both waveguides.

\subsection{Asymmetric coupled waveguides: LC symmetry}

Although $\mathcal{PT}$-symmetric systems allow perfect loss compensation, it is often problematic to reach the ideal symmetry and the ideal loss-gain balance in realistic situations. Here, we show that loss compensation can be obtained in asymmetric systems with unequal coupled waveguides. Such situation can be described as LC symmetry \cite{Klimov_LaserPhysLett_2018}, which has much in common with $\mathcal{PT}$ symmetry (e.g., the existence of EPs), but at the same time, the requirements for system symmetry are strongly relaxed. Moreover, LC symmetry is reached for a pair of different modes as will be illustrated further.

\begin{figure}
\includegraphics[width=\linewidth]{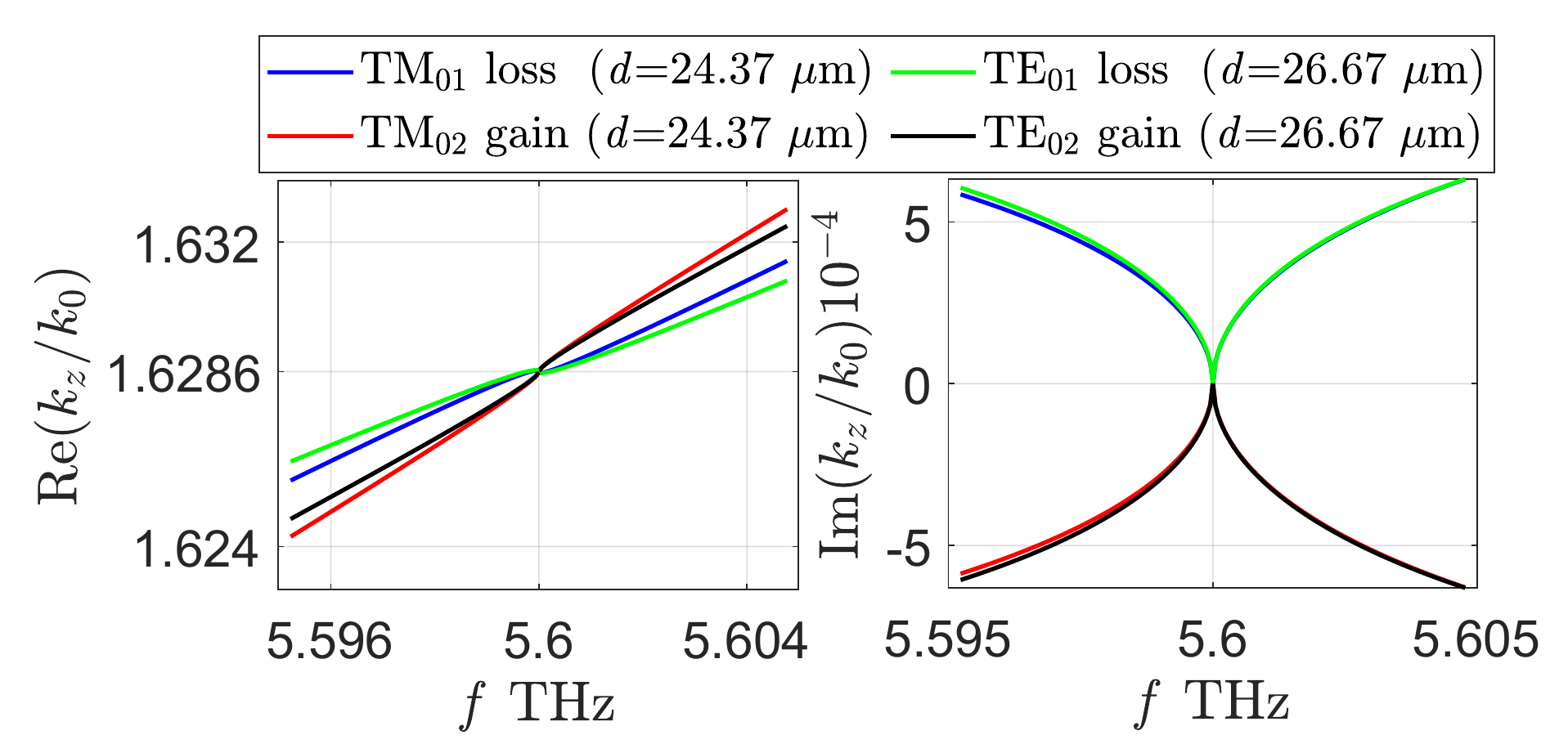}
\caption{\label{disp_tm_te} EPs corresponding to the LC symmetry for the pairs of modes TM$_{01}$-TM$_{02}$ and TE$_{01}$-TE$_{02}$.}
\end{figure}

Let us take the loss tangent of the first cylinder equal to $\tan{\delta_1}=5\cdot10^{-4}$ corresponding to the value for silicon \cite{Lamb_Inf_Mill_Wave_1996}. Changing the gain tangent $\tan{\delta_2}$ and the distance between the cylinders $d$, we can obtain the full loss compensation at the target frequency $f=5.6$~THz and wavenumber $k_z/k_0=1.6286$. For the TE$_{01}$ and TE$_{02}$ modes, the EP which we call the LC-symmetry threshold is reached at the system parameters as follows: $R_1=7.48$~$\mu$m, $R_2=16.305$~$\mu$m, $\epsilon_{r1}=\epsilon_{r2}=12$, $d=26.67$~$\mu$m, $\tan{\delta_2}=-4.05\cdot10^{-4}$) (see Fig. \ref{disp_tm_te}, green and black lines). For the modes TM$_{01}$ and TM$_{02}$, the EP is reached at $d=24.37$~$\mu$m and $\tan{\delta_2}=-4.05\cdot10^{-4}$ (see Fig. \ref{disp_tm_te}, blue and red lines). Similarly to the $\mathcal{PT}$-symmetry threshold, the LC-symmetry threshold corresponds to the purely real eigenvalues $k_z/k_0$. This approach can be utilized to obtain the EPs of full loss compensations for other pairs of modes as well.

\begin{figure}
\includegraphics[width=\linewidth]{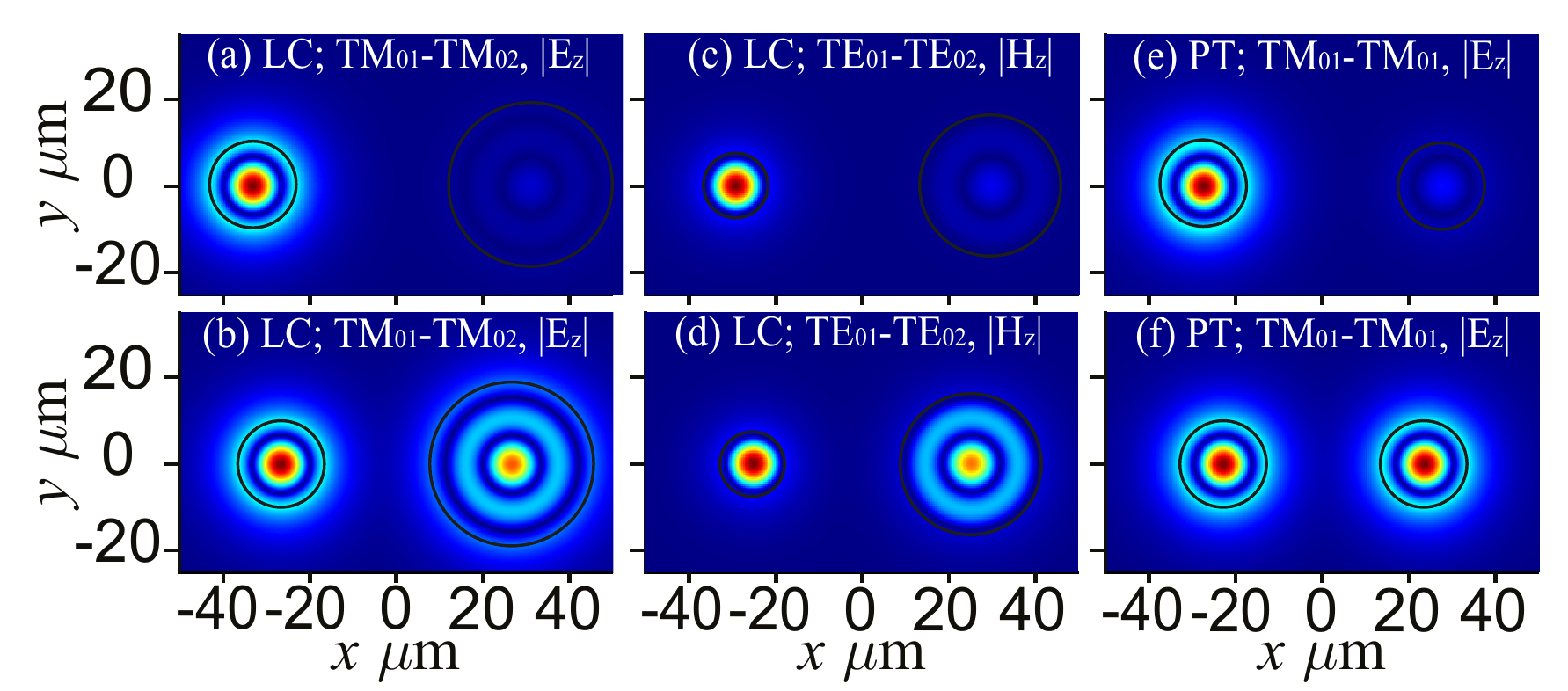}
\caption{\label{fig_field_tm_te} (a), (c) Electric $|E_z|$ and Magnetic $|H_z|$  field components corresponding to the broken LC-symmetry case for TM$_{01}$ and TE$_{01}$ loss modes at distance $d=35~\mu$m.  (b), (d)  Electric $|E_z|$ and Magnetic $|H_z|$  field components corresponding to the LC-symmetry thresholds of TM$_{01}$-TM$_{02}$ and TE$_{01}$-TE$_{02}$ mode pairs at distances $d=24.37~\mu$m and $d=26.67~\mu$m. (e), (f) Electric field $|E_z|$ corresponding to the asymmetry and $\mathcal{PT}$-symmetry cases of TM$_{01}$-TM$_{01}$ mode pairs at distances $d=35~\mu$m and $d=26.31~\mu$m.}
\end{figure}
% (a), (e) Electric field $|E_z|$ and (c) Magnetic field $|H_z|$ 
In Figs. \ref{fig_coupling_tm} and \ref{fig_coupling_te}, the LC-symmetry thresholds are shown for the TM-type and TE-type modes, respectively, and compared to the $\mathcal{PT}$-symmetry thresholds. One can see that the LC-symmetry thresholds for the different modes (for example, TM$_{01}$ and TM$_{02}$) lie between the $\mathcal{PT}$-symmetry thresholds for the corresponding individual modes. The main difference with the $\mathcal{PT}$-symmetric case is that the eigenvalues for the asymmetric waveguides do not remain real below the LC-symmetry threshold. In fact, full loss compensation is realized only in a single point. The position of this point -- LC-symmetry threshold -- depends on the loss tangent and the frequency at which the dispersion curves cross. An example of electric and magnetic field distributions at the LC-symmetry threshold and dots of broken symmetry for dissimilar waveguides are shown in Figs. \ref{fig_field_tm_te}(b) and \ref{fig_field_tm_te}(d) and Figs. \ref{fig_field_tm_te}(a) and \ref{fig_field_tm_te}(c) as compared to the distributions at the $\mathcal{PT}$-symmetry thresholds and case of broken $\mathcal{PT}$-symmetry for the identical waveguides in Fig. \ref{fig_field_tm_te}(f) and \ref{fig_field_tm_te}(e).  The smaller the loss tangent and the nearer the crossing to the cutoff frequency, the larger the distance between the waveguides corresponding to the threshold \cite{Hlushchenko_LaserPhysLett_2020}. 

Thus, the LC-symmetry threshold in the asymmetric system allows one to fully compensate the loss for the unbalanced gain and loss of the individual cylinders. This effect can be considered as a generalization of $\mathcal{PT}$-symmetric loss compensation, since both the dispersion curves and the behavior of electromagnetic fields show that the nature of loss compensation phenomenon in asymmetric systems is similar to the $\mathcal{PT}$ symmetry in symmetric ones.

\section{Hybrid modes: LC-symmetry}

\subsection{Modes with the same azimuthal index}

We have shown above how LC and $\mathcal{PT}$ symmetries can be observed by coupling either TM or TE modes of the pair of cylindrical waveguides. However, when one mode is TM one and another is TE one, we cannot obtain an EP due to the weak coupling between the modes of different types (see Fig. \ref{fig_coupling_te}, green and pink lines). In order to get around this problem, in this section, we consider the possibility of LC symmetry for the hybrid modes, which can be treated as a linear superposition of the corresponding TE and TM modes. These modes having both the electric and magnetic field components in the longitudinal direction can be either of HE type (magnetic component dominates) or EH type (electric component dominates).

\begin{figure}
\includegraphics[width=\linewidth]{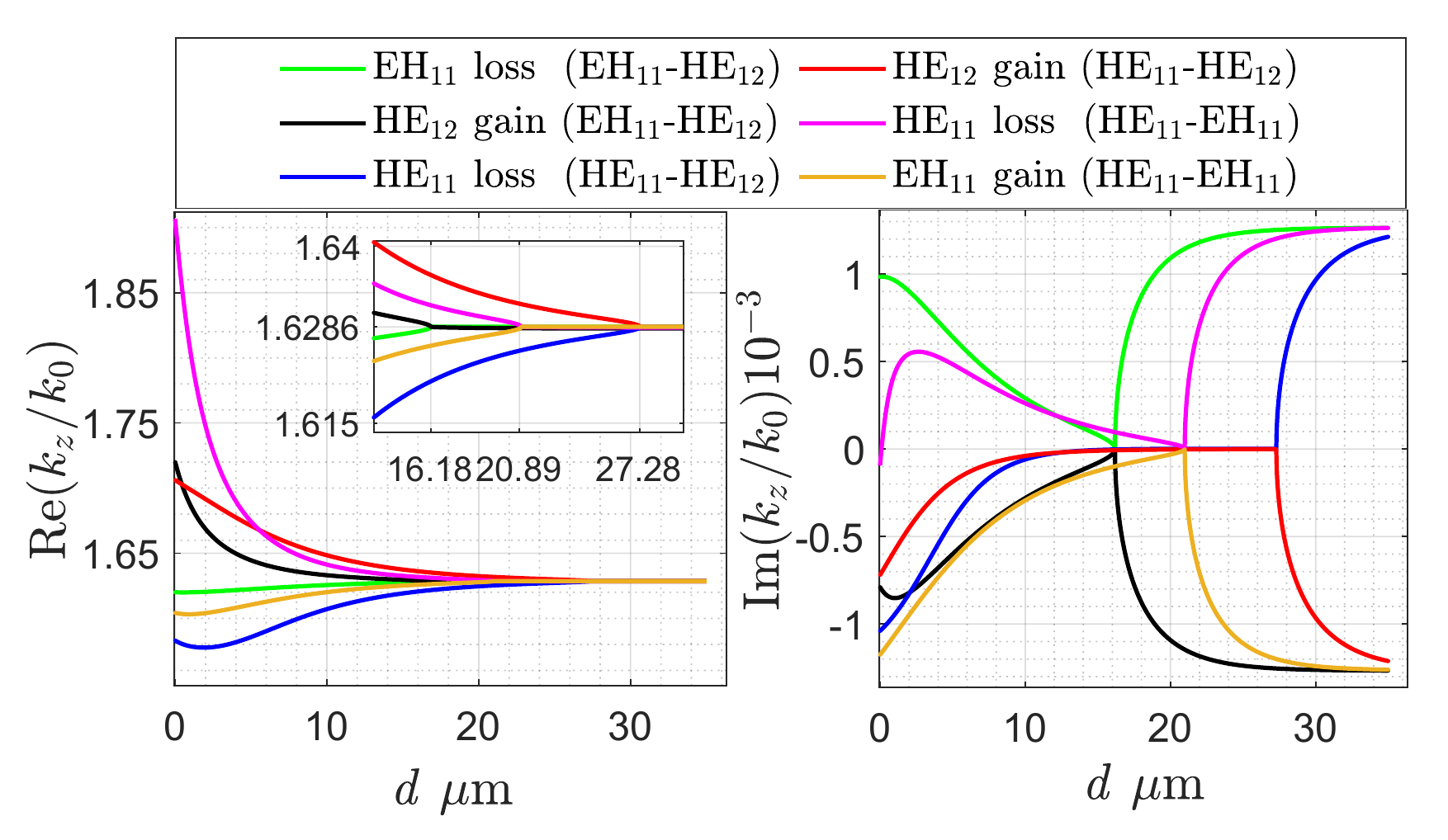}
\caption{\label{fig_hybrid_1} LC symmetry for the hybrid modes with the azimuthal index $1$.}
\includegraphics[width=\linewidth]{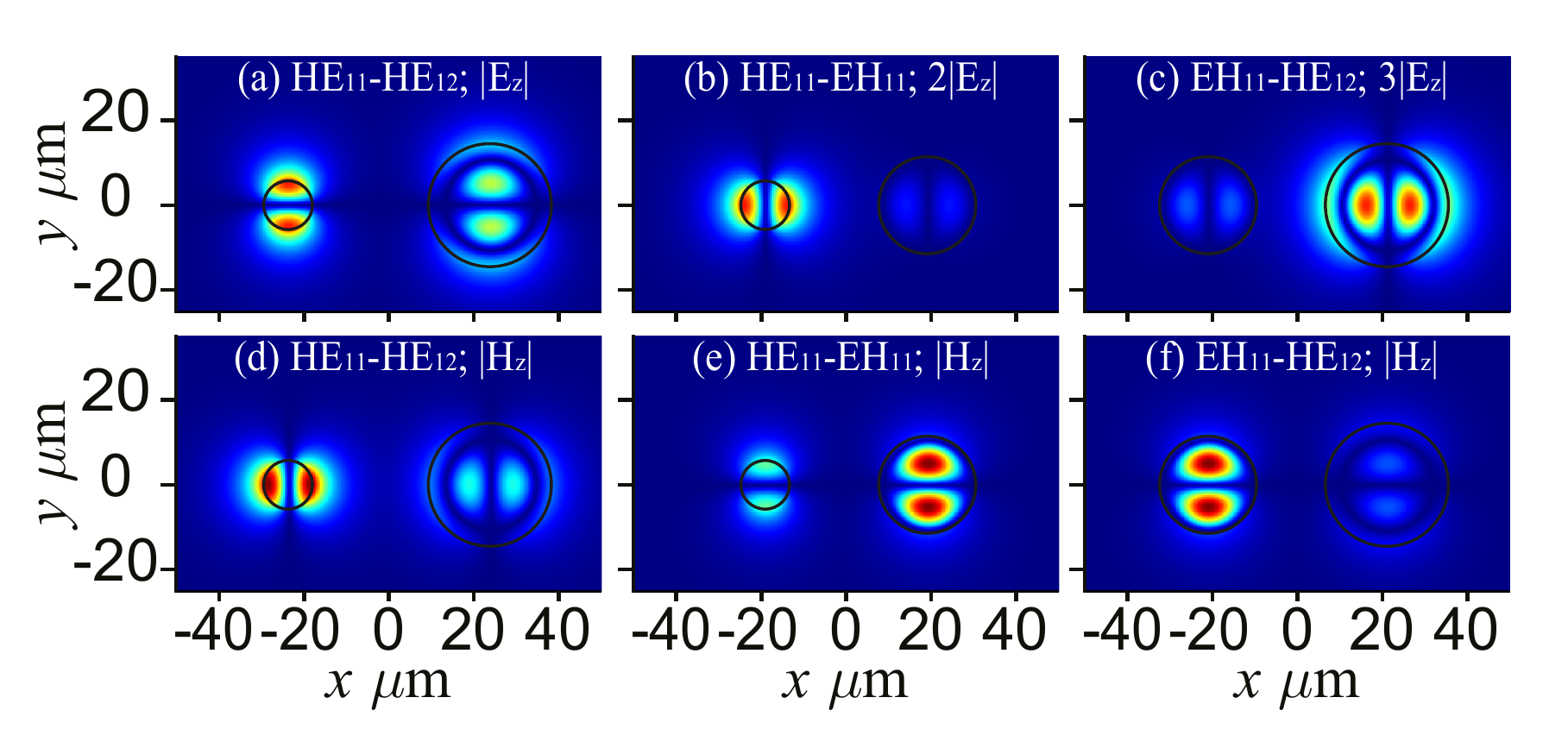}
\caption{\label{fig_field_hybrid_1} (a), (b), (c) Electric field $|E_z|$ corresponding to the LC-symmetry thresholds of HE$_{11}$-HE$_{12}$, HE$_{11}$-EH$_{11}$ and EH$_{11}$-HE$_{12}$ mode pairs; (d), (e), (f) The corresponding magnetic field $|H_z|$.}
\includegraphics[width=\linewidth]{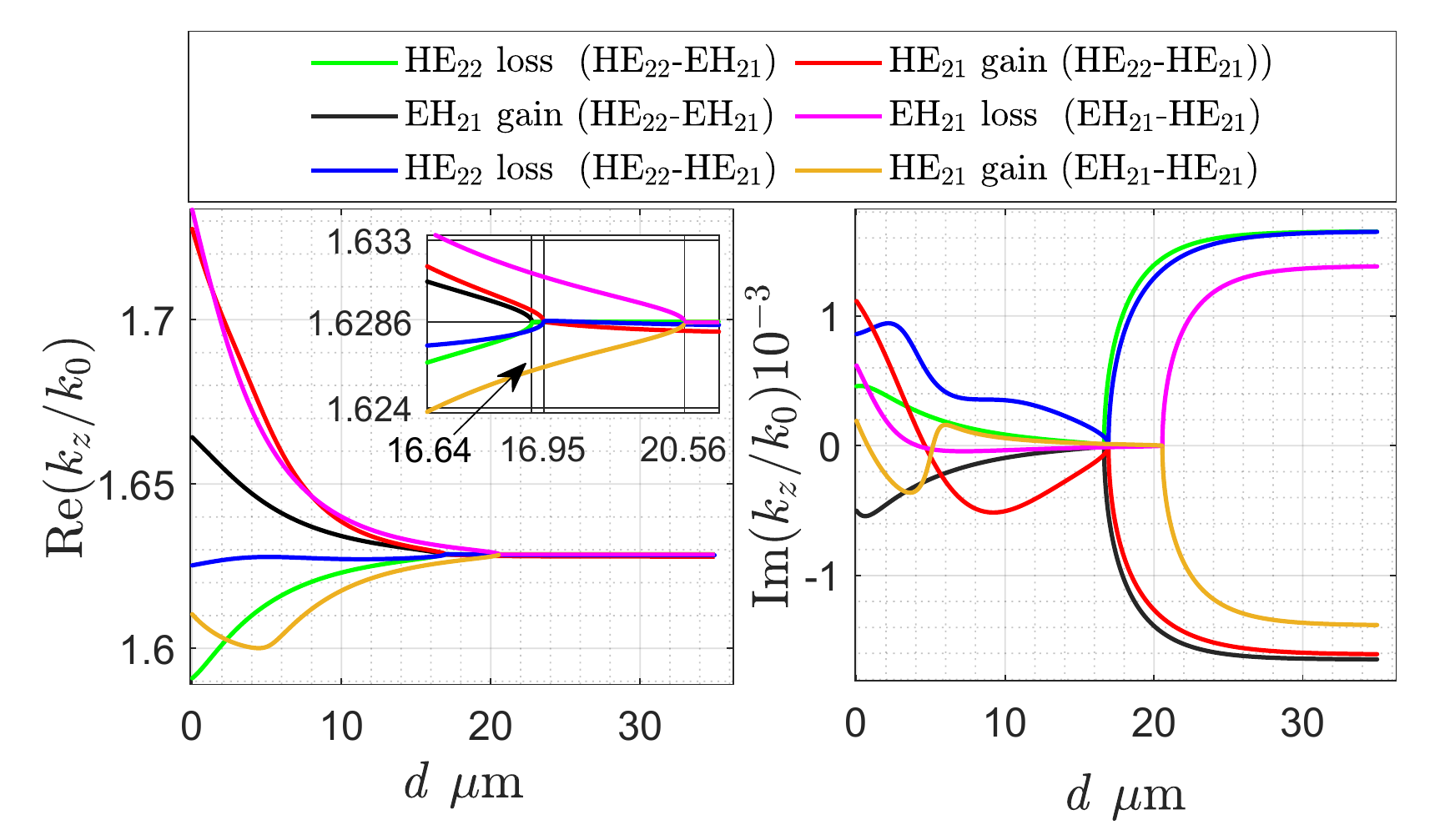}
\caption{\label{fig_hybrid_2} LC-symmetry of hybrid modes with the azimuthal index $2$.}
\includegraphics[width=\linewidth]{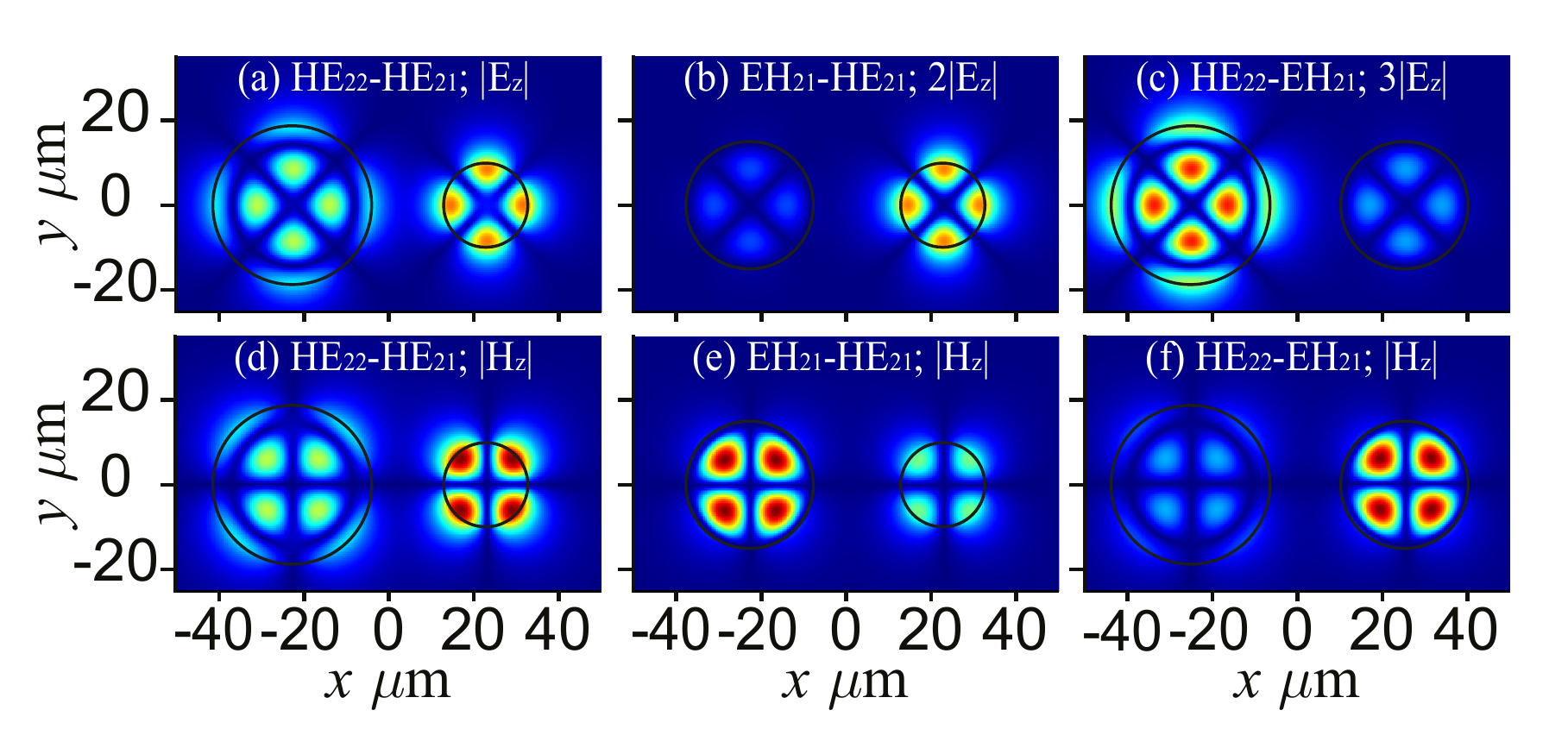}
\caption{\label{fig_field_hybrid_2} (a), (b), (c) Electric field $|E_z|$ corresponding to the LC-symmetry thresholds of HE$_{22}$-HE$_{21}$, EH$_{21}$-HE$_{21}$ and HE$_{22}$-EH$_{21}$ mode pairs; (d), (e), (f) The corresponding magnetic field $|H_z|$.}
\end{figure}

We tune the dispersion curves for two cylinders to obtain the LC-symmetry thresholds of hybrid modes for the same parameters ($f=5.6$~THz and $k_z/k_0=1.6286$) as for the TM and TE modes. As previously, we start from the modes of the individual waveguide and shift the dispersion curves by changing the radius to get the required mode at the target frequency and wavenumber. For example, we obtain the hybrid mode HE$_{11}$ at $f=5.6$~THz and $k_z/k_0=1.6286$ for the cylinder with $R=5.732$~$\mu$m (see Fig. \ref{fig_full_modes}, dashed line).

Let us consider the class of hybrid modes with azimuthal index $1$ focusing in particular on the modes HE$_{11}$, EH$_{11}$ and HE$_{12}$ (see Fig. \ref{fig_full_modes}, black lines). We are interested in the LC symmetry between the mode pairs HE$_{11}$-EH$_{11}$, HE$_{11}$-HE$_{12}$ and EH$_{11}$-HE$_{12}$ with the first mode corresponding to the cylinder with loss (Region I) and the second mode corresponding to the cylinder with gain (Region II). As shown in Fig. \ref{fig_hybrid_1}, the LC-symmetry thresholds exist for all the mode pairs mentioned above and for the system parameters as follows:

\begin{itemize}
  \item HE$_{11}$-EH$_{11}$ -- \{$R_1=5.732$~$\mu$m, $R_2=11.44$~$\mu$m, $\epsilon_{r1}=\epsilon_{r2}=12$, $\tan{\delta_1}=5\cdot10^{-4}$, $\tan{\delta_2}=-4.91\cdot10^{-4}$\};
  \item HE$_{11}$-HE$_{12}$ -- \{$R_1=5.732$~$\mu$m, $R_2=14.487$~$\mu$m, $\epsilon_{r1}=\epsilon_{r2}=12$, $\tan{\delta_1}=5\cdot10^{-4}$, $\tan{\delta_2}=-4.33\cdot10^{-4}$\};
  \item EH$_{11}$-HE$_{12}$ -- \{$R_1=11.44$~$\mu$m, $R_2=14.487$~$\mu$m, $\epsilon_{r1}=\epsilon_{r2}=12$, $\tan{\delta_1}=4.91\cdot10^{-4}$, $\tan{\delta_2}=-4.33\cdot10^{-4}$\}.
\end{itemize}

We see that the LC-symmetry thresholds are reached for different distances between waveguides for every pair of modes: $d=20.89$~$\mu$m for HE$_{11}$-EH$_{11}$; $d=27.28$~$\mu$m for HE$_{11}$-HE$_{12}$; and $d=16.18$~$\mu$m for EH$_{11}$-HE$_{12}$. The field profiles shown in Fig. \ref{fig_field_hybrid_1} prove that there are nonzero $E_z$ and $H_z$ in both cylinders as expected for the hybrid modes (although contribution of electric or magnetic field can strongly differ).

As to the hybrid modes with the azimuthal index $2$, we choose the mode pairs as follows: HE$_{22}$-EH$_{21}$, HE$_{22}$-HE$_{21}$ and EH$_{21}$-HE$_{21}$. Now, unlike previous cases, the radius of the waveguide with loss should be taken larger than the radius of the cylinder with gain. For the parameters of the corresponding LC-symmetry thresholds, shown in Figs. \ref{fig_hybrid_2} and \ref{fig_field_hybrid_2}, we have:

\begin{itemize}
  \item HE$_{22}$-EH$_{21}$ -- \{$R_1=18.7$~$\mu$m, $R_2=15$~$\mu$m, $\epsilon_{r1}=\epsilon_{r2}=12$, $\tan{\delta_1}=5\cdot10^{-4}$, $\tan{\delta_2}=-5.95\cdot10^{-4}$\};
  \item HE$_{22}$-HE$_{21}$ -- \{$R_1=18.7$~$\mu$m, $R_2=9.91$~$\mu$m, $\epsilon_{r1}=\epsilon_{r2}=12$, $\tan{(\delta_1)}=5\cdot10^{-4}$, $\tan{(\delta_2)}=-4.65\cdot10^{-4}$\};
  \item EH$_{21}$-HE$_{21}$ -- \{$R_1=15$~$\mu$m, $R_2=9.91$~$\mu$m, $\epsilon_{r1}=\epsilon_{r2}=12$, $\tan{\delta_1}=5\cdot10^{-4}$, $\tan{\delta_2}=-4\cdot10^{-4}$\}.
\end{itemize}

The distances between the waveguides corresponding to these thresholds are: $d=16.64$~$\mu$m for HE$_{22}$-EH$_{21}$;  $d=16.95$~$\mu$m for HE$_{22}$-HE$_{21}$; and $d=20.56$~$\mu$m for EH$_{21}$-HE$_{21}$. The results in Figs.  \ref{fig_hybrid_1}-\ref{fig_field_hybrid_2} evidence that the full loss compensation can be realized for the arbitrary hybrid modes with the same azimuthal index. The structure of the modes influence only which of the cylinders should have larger radius -- the lossy or the gainy one.

\subsection{Modes with different azimuthal indices}

In this subsection, we study the possibility of LC symmetry in the coupled waveguides with loss and gain tuned to the modes with different azimuthal indices. The fundamental difference to the cases considered above is the necessity to take into account the larger number of azimuthal harmonics. For the identical azimuthal indices from the previous subsection, one can obtain the LC-symmetry threshold with just one harmonic, i.e., leaving a single term in the sums of Eqs. (\ref{field_1})-(\ref{field_3}). On the contrary, for the modes with different azimuthal indices, several harmonics are needed due to weak coupling between the fields (similar to the TM-TE pair).

\begin{figure}
\includegraphics[width=\linewidth]{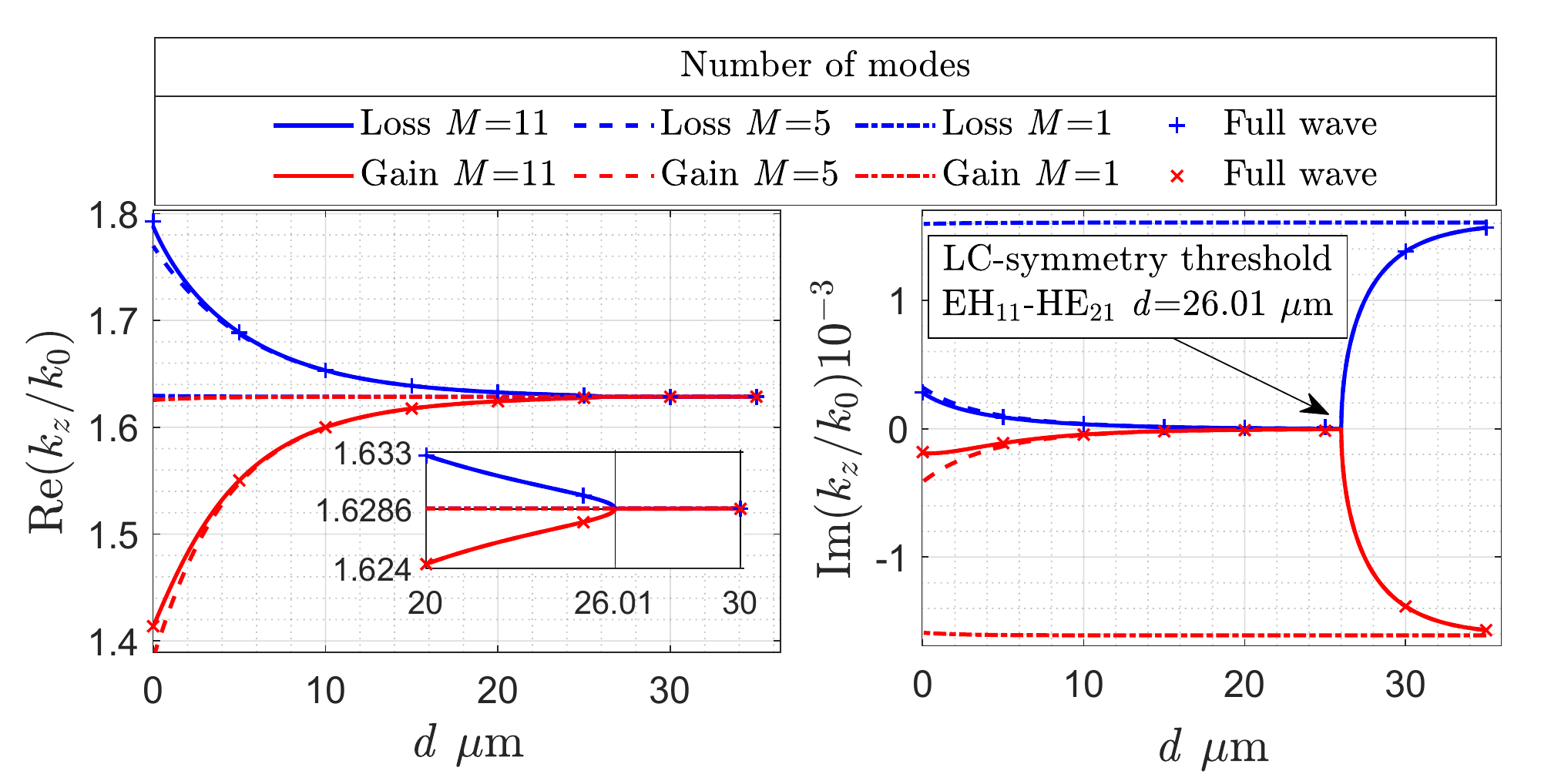}
\caption{\label{fig_hybrid_1_2} LC symmetry for the hybrid modes EH$_{11}$-HE$_{21}$ having different azimuthal indices.}
\includegraphics[width=\linewidth]{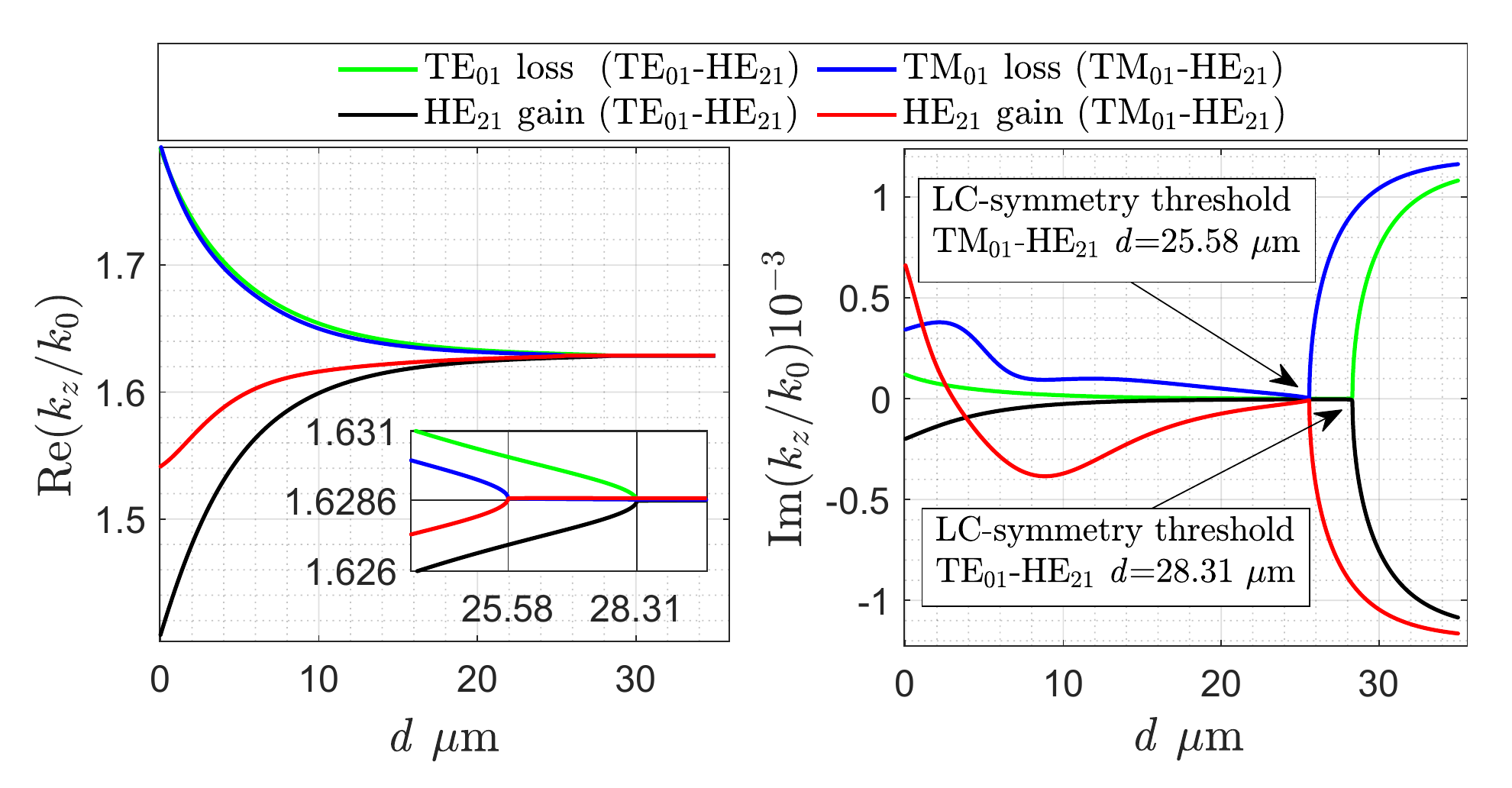}
\caption{\label{fig_hybrid_0_2} LC symmetry for the circularly symmetric TM$_{01}$, TE$_{01}$ and hybrid HE$_{21}$ modes.}
\includegraphics[width=\linewidth]{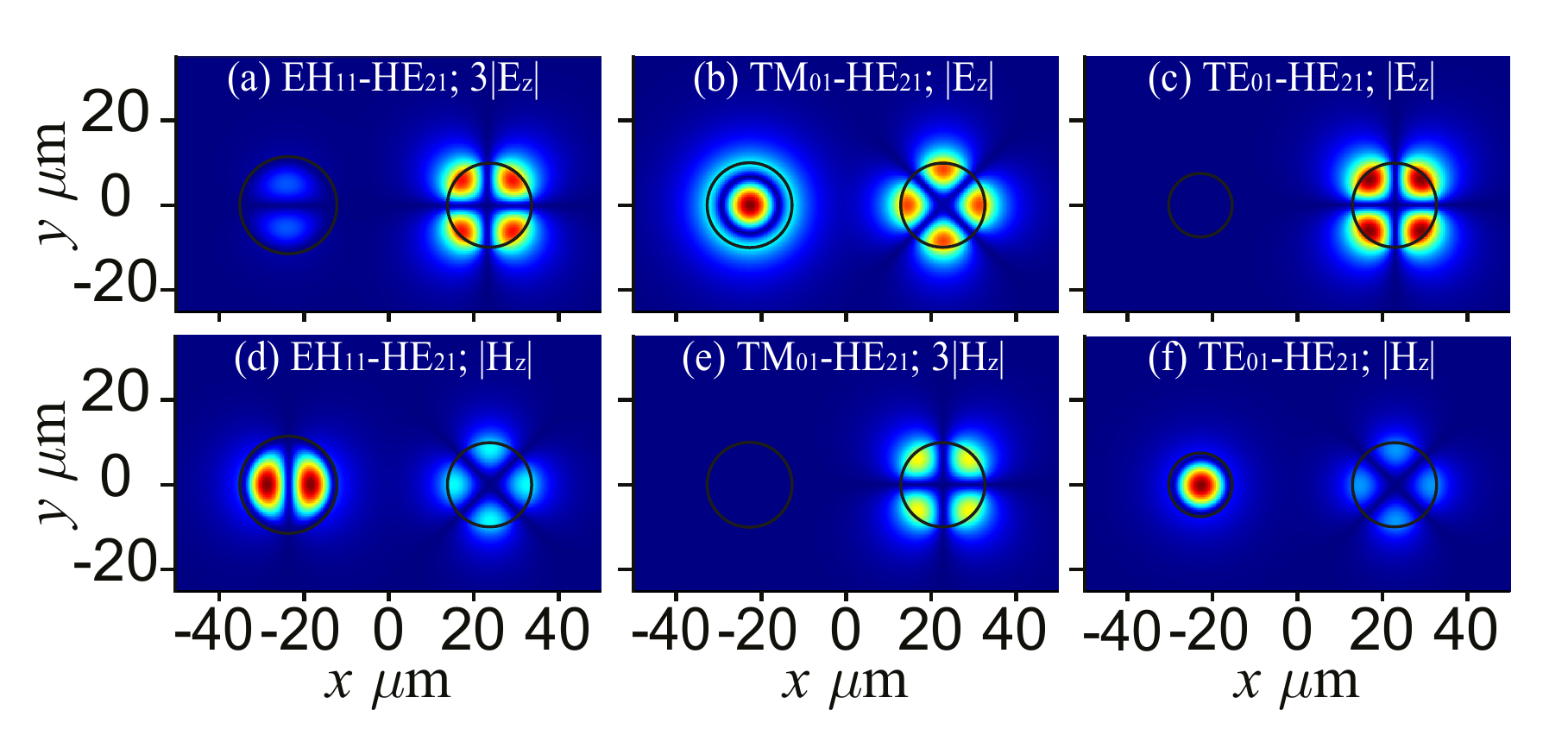}
\caption{\label{fig_field_hybrid_0_1_2} (a), (b), (c) Electric field $|E_z|$ corresponding to the LC-symmetry thresholds of EH$_{11}$-HE$_{21}$, TM$_{01}$-HE$_{21}$ and TE$_{01}$-HE$_{21}$ mode pairs; (d), (e), (f) The corresponding magnetic field $|H_z|$.}
\end{figure}

As an example, let us consider the hybrid modes HE$_{21}$ and EH$_{11}$. For their dispersion curves to cross at $f=5.6$~THz and $k_z/k_0=1.6286$, the radii of the cylinders should be $R_1=11.44$~$\mu$m and $R_2=9.91$~$\mu$m. Figure \ref{fig_hybrid_1_2} shows how the number of azimuthal harmonics in Eqs. (\ref{field_1})-(\ref{field_3}) influences the possibility of the loss compensation phenomenon for these modes. In this figure, $M=2N+1$ is the number of terms took into account in the field sums. We see that the LC-symmetry threshold is observed for the values $\tan{\delta_1}=6.23\cdot10^{-4}$ and $\tan{\delta_2}=-4.65\cdot10^{-4}$ at the distance $d=26.01$~$\mu$m between the waveguides only for $M \geq 5$, i.e., the minimal set of harmonics includes the terms with $n=\{0,\pm1, \pm2\}$. This result is confirmed by the comparison with the independent full-wave calculation using COMSOL Multiphysics{\textsuperscript{\tiny\textregistered}} software (see symbols in Fig. \ref{fig_hybrid_1_2}). The required number of harmonics becomes larger for very close cylinders (near $d=0$) in order to save the consistency between the analytical and numerical methods as demonstrated with the calculations for $M=11$ in Fig. \ref{fig_hybrid_1_2}. Further, we omit the comparison with the full-wave simulations, since we are interested mainly in the region close to the LC-symmetry thresholds where the consistency is perfect even for relatively small $M$.

Since the hybrid modes can be coupled to the symmetric TM and TE modes, we analyze the loss compensation for such situations as well. For the pair TM$_{01}$-HE$_{21}$, for example, the LC-symmetry threshold can be reached at $d=25.58$~$\mu$m, whereas it is $d=28.31$~$\mu$m for TE$_{01}$-HE$_{21}$ (see Fig. \ref{fig_hybrid_0_2}). The parameters of the system should be taken as follows:

\begin{itemize}
\item  TM$_{01}$-HE$_{21}$ -- \{$R_1=10$~$\mu$m, $R_2=9.91$~$\mu$m, $\epsilon_{r1}=\epsilon_{r2}=12$, $\tan{\delta_1}=5\cdot10^{-4}$, $\tan{\delta_2}=-3.44\cdot10^{-4}$\};
\item TE$_{01}$-HE$_{21}$ -- \{$R_1=7.48$~$\mu$m, $R_2=9.91$~$\mu$m, $\epsilon_{r1}=\epsilon_{r2}=12$, $\tan{\delta_1}=5\cdot10^{-4}$, $\tan{\delta_2}=-3.32\cdot10^{-4}$\};
\end{itemize}

Note that for the TM$_{01}$-HE$_{21}$ pair at the LC-symmetry threshold, the longitudinal electric field $E_z$ is excited in both cylinders [Fig. \ref{fig_field_hybrid_0_1_2}(b)], whereas the longitudinal magnetic field $H_z$ exists only in the second one due to the HE$_{21}$ mode [Fig. \ref{fig_field_hybrid_0_1_2}(e)]. The opposite is true for the TE$_{01}$-HE$_{21}$ pair: $E_z$ is absent in the first cylinder [Fig. \ref{fig_field_hybrid_0_1_2}(c)], but $H_z$ is present in both waveguides [Fig. \ref{fig_field_hybrid_0_1_2}(f)].

\section{Conclusion}

We employ the multi-mode analytical approach to study the $\mathcal{PT}$ and LC symmetries in the dielectric cylindrical waveguides with gain and loss. It is shown that for any predetermined frequency, one can realize the loss compensation by tuning the waveguides to the proper modes and to the proper radii. For the modes of either TM or TE nature, we can reach loss compensation either in the $\mathcal{PT}$-symmetric phase (when the waveguides have identical radii) or at the LC-symmetry threshold (when the waveguides are dissimilar). The loss compensation is also possible in the situations involving the hybrid modes with identical or different azimuthal indices. By tuning the waveguides radii, we obtained the LC-symmetry thresholds for a wide number of mode pairs, namely \{HE-HE, HE-EH, HE-TM, HE-TE\}. The results are illustrated with the profiles of $E_z$ and $H_z$ components of electromagnetic field at the corresponding EPs (LC- and $\mathcal{PT}$-symmetry thresholds) and corroborated by comparing with the full-wave simulations.

Our results show that the loss compensation phenomenon in loss-gain systems is the general effect taking place even in asymmetric systems, such as coupled dielectric waveguides of different radii. Tuning the waveguides to the desired modes simply by changing the distance between them and their radii allows to build the lossless system starting from the arbitrary values of gain and loss. These results significantly expand the understanding of the concepts of LC and $\mathcal{PT}$ symmetry and can be applied for the experimental implementation of new types of optical devices.

\acknowledgements{The work was supported by the National Key R\&D Program of China (Project No. 2018YFE0119900) and the State Committee on Science and Technology of Belarus (Project No. F20KITG-010).}

\begin{widetext}

\section*{Appendix: Derivation of transverse field components}\label{Appendix}

The transverse field components $(E_r, H_r, E_{\phi}, H_{\phi} )$ in all regions can be readily expressed in terms of axial components Eqs.~(\ref{field_1})-(\ref{field_3}) from the Maxwell equations.
The $\phi$-components of the fields in the first cylinder (Region I) can be derived as:
\begin{equation}\label{phi_1}
\begin{split}
    E_{\phi}^1&=\frac{1}{k_{p,1}^2}\sum_n{\left[-\frac{k_zn}{r_1}A_n^1J_n(k_{p,1}r_1)-i\omega\mu_0B_n^1k_{p,1}J'_n(k_{p,1}r_1)\right]e^{in\phi_1}},\\
    H_{\phi}^1&=\frac{1}{k_{p,1}^2}\sum_n{\left[i\omega\epsilon_1k_{p,q}A_n^1J'_n(k_{p,1}r_1)-\frac{k_zn}{r_1}B_n^1J_n(k_{p,1}r_1)\right]e^{in\phi_1}},
\end{split}
\end{equation}
The $\phi$-components of the fields inside the second cylinder (Region II):
\begin{equation}\label{phi_2}
\begin{split}
    E_{\phi}^2&=\frac{1}{k_{p,2}^2}\sum_n{\left[-\frac{k_zn}{r_2}A_n^2J_n(k_{p,2}r_2)-i\omega\mu_0B_n^2k_{p,q}J'_n(k_{p,2}r_2)\right]e^{in\phi_2}},\\
    H_{\phi}^2&=\frac{1}{k_{p,2}^2}\sum_n{\left[i\omega\epsilon_2k_{p,2}A_n^2J'_n(k_{p,2}r_2)-\frac{k_zn}{r_2}B_n^2J_n(k_{p,2}r_2)\right]e^{in\phi_2}},
\end{split}
\end{equation}

For derivation of the fields in the Region III, we connect the coordinates $(r_1,\phi_1)$ and $(r_2,\phi_2)$ using the Graf addition theorem:
\begin{equation}\label{graf_th}
\begin{split}
B_n(r_1)e^{\pm in\phi_1}=\sum_{k=-N}^{N}{B_{n+k}(h)J_k(r_2)e^{\mp i k\phi_2}}e ^{\pm i k\pi}, \\
B_n(r_2)e^{\pm in\phi_2}=\sum_{k=-N}^{N}{B_{n+k}(h)J_k(r_1)e^{\mp i k\phi_1}}e ^{\pm i n\pi},
\end{split}
\end{equation}
where $B_n(\cdot)$ is the $n$-th order cylindrical function and $h$ is the distance between the centers of two cylinders ($h=d+R_1+R_2$) (see Fig. \ref{fig_cylinders}). The $\phi$-components of the fields outside the cylinders in coordinates $(r_1,\phi_1)$ are:
\begin{equation}\label{phi_3_r1}
\begin{split}
   E_{\phi}^3(r_1,\phi_1)=&\frac{1}{k_{p,3}^2}\sum_n\left\{-\frac{k_zn}{r_1}\left[C_n^1H_n^{(1,2)}(k_{p,3}r_1)e^{in\phi_1}+\sum_k{C_n^2H_{n+k}^{(1,2)}(k_{p,3}h)J_k(k_{p,3}r_1)e^{-ik\phi_1}e^{in\pi}}\right]-\right.\\ &\left.-i\omega\mu_0B_n^2k_{p,3} \left[D_n^1H_n^{'(1,2)}(k_{p,3}r_1)e^{in\phi_1}+\sum_k{D_n^2H_{n+k}^{(1,2)}(k_{p,3}h)J'_k(k_{p,3}r_1)e^{-ik\phi_1}e^{in\pi}}\right]\right\},\\
  H_{\phi}^3(r_1,\phi_1)=&\frac{1}{k_{p,3}^2}\sum_n\left\{i\omega\epsilon_3k_{p,3}\left[C_n^1H_n^{'(1,2)}(k_{p,3}r_1)e^{in\phi_1}+\sum_k{C_n^2H_{n+k}^{(1,2)}(k_{p,3}h)J'_k(k_{p,3}r_1)e^{-ik\phi_1}e^{in\pi}}\right]\right.-\\
  &-\left.\frac{k_zn}{r_1}\left[D_n^1H_n^{(1,2)}(k_{p,3}r_1)e^{in\phi_1}+\sum_k{D_n^2H_{n+k}^{(1,2)}(k_{p,3}h)J_k(k_{p,3}r_1)e^{-ik\phi_1}e^{in\pi}}\right]\right\},
\end{split}
\end{equation}
The $\phi$-components of the  fields in the Region III in the coordinates $(r_2,\phi_2)$ are:
\begin{equation}\label{phi_3_r2}
\begin{split}
   E_{\phi}^3(r_2,\phi_2)=&\frac{1}{k_{p,3}^2}\sum_n\left\{-\frac{k_zn}{r_2}\left[\sum_k{C_n^1H_{n+k}^{(1,2)}(k_{p,3}h)J_k(k_{p,3}r_2)e^{ik(\pi-\phi_2)}}+C_n^2H_n^{(1,2)}(k_{p,3}r_2)e^{in\phi_2}\right]-\right.\\ &\left.-i\omega\mu_0k_{p,3} \left[\sum_k{D_n^1H_{n+k}^{(1,2)}(k_{p,3}h)J'_k(k_{p,3}r_2)e^{ik(\pi-\phi_2)}}+D_n^2H_n^{'(1,2)}(k_{p,3}r_2)e^{in\phi_2}\right]\right\},\\
 H_{\phi}^3(r_2,\phi_2)=&\frac{1}{k_{p,3}^2}\sum_n\left\{i\omega\epsilon_3k_{p,3}\left[\sum_k{C_n^1H_{n+k}^{(1,2)}(k_{p,3}h)J'_k(k_{p,3}r_2)e^{ik(\pi-\phi_2)}}+C_n^2H_n^{'(1,2)}(k_{p,3}r_2)e^{in\phi_2}\right]\right.-\\
  &-\left.\frac{k_zn}{r_2}\left[\sum_k{D_n^1H_{n+k}^{(1,2)}(k_{p,3}h)J_k(k_{p,3}r_2)e^{ik(\pi-\phi_2)}}+D_n^2H_n^{(1,2)}(k_{p,3}r_2)e^{in\phi_2}\right]\right\},
\end{split}
\end{equation}

The explicit form of the dispersion relation is the condition of zero determinant of the following system of equations:
\begin{equation}\label{boun_cond_full}
\begin{split}
\tilde{A}_m^1J_m(k_{p,1}r_1)=\tilde{C}_m^1H_m^{(1,2)}(k_{p,3}r_1)+\sum_n{\tilde{C}_n^2H_{m-n}^{(1,2)}(k_{p,3}h)J_m(k_{p,3}r_1)},\\
B_m^1J_m(k_{p,1}r_1)=D_m^1H_m^{(1,2)}(k_{p,3}r_1)+\sum_n{D_n^2H_{m-n}^{(1,2)}(k_{p,3}h)J_m(k_{p,3}r_1)},\\
\tilde{A}_m^2J_m(k_{p,2}r_2)=\sum_n{\tilde{C}_n^1H_{m-n}^{(1,2)}(k_{p,3}h)J_m(k_{p,3}r_2)}+\tilde{C}_m^2H_m^{(1,2)}(k_{p,3}r_2),\\
B_m^2J_m(k_{p,2}r_2)=\sum_n{D_n^1H_{m-n}^{(1,2)}(k_{p,3}h)J_m(k_{p,3}r_2)}+D_m^2H_m^{(1,2)}(k_{p,3}r_2), \\
\frac{1}{k_{p,1}^2}\left[-\frac{k_zm}{r_1}\tilde{A}_m^1J_m(k_{p,1}r_1)-ik_0B_m^1k_{p,1}J'_m(k_{p,1}r_1)\right]=\\
=\frac{1}{k_{p,3}^2}\left\{-\frac{k_zm}{r_1}\left[\tilde{C}_m^1H_m^{(1,2)}(k_{p,3}r_1)+\sum_n{\tilde{C}_n^2H_{m-n}^{(1,2)}(k_{p,3}h)J_m(k_{p,3}r_1)}\right]\right.-\\
-\left.ik_0k_{p,3}\left[D_m^1H_m^{'(1,2)}(k_{p,3}r_1)+\sum_n{D_n^2}H_{m-n}^{(1,2)}(k_{p,3}h)J'_m(k_{p,3}r_1)\right]\right\},\\
\frac{1}{k_{p,1}^2}\left[ik_0\epsilon_1k_{p,1}\tilde{A}_m^1J'_m(k_{p,1}r_1)-\frac{k_zm}{r_1}B_m^1J_m(k_{p,1}r_1)\right]=\\
=\frac{1}{k_{p,3}^2}\left\{ik_0\epsilon_3k_{p,3}\left[\tilde{C}_m^1H_m^{'(1,2)}(k_{p,3}r_1)+\sum_n{\tilde{C}_n^2H_{m-n}^{(1,2)}(k_{p,3}h)J'_m(k_{p,3}r_1)}\right]\right.-\\
-\left.\frac{k_zm}{r_1}\left[D_m^1H_m^{(1,2)}(k_{p,3}r_1)+\sum_n{D_n^2}H_{m-n}^{(1,2)}(k_{p,3}h)J_m(k_{p,3}r_1)\right]\right\},\\
\frac{1}{k_{p,2}^2}\left[-\frac{k_zm}{r_2}\tilde{A}_m^2J_m(k_{p,2}r_2)-ik_0B_m^2k_{p,2}J'_m(k_{p,2}r_2)\right]=\\
=\frac{1}{k_{p,3}^2}\left\{-\frac{k_zm}{r_2}\left[\sum_n{\tilde{C}_n^1H_{m-n}^{(1,2)}(k_{p,3}h)J_m(k_{p,3}r_2)}+\tilde{C}_m^2H_m^{(1,2)}(k_{p,3}r_2)\right]\right.-\\
-\left.ik_0k_{p,3}\left[\sum_n{D_n^1}H_{m-n}^{(1,2)}(k_{p,3}h)J'_m(k_{p,3}r_2)+D_m^2H_m^{'(1,2)}(k_{p,3}r_2)\right]\right\},\\
\frac{1}{k_{p,2}^2}\left[ik_0\epsilon_2k_{p,2}\tilde{A}_m^2J'_m(k_{p,2}r_2)-\frac{k_zm}{r_2}B_m^2J_m(k_{p,2}r_2)\right]=\\
=\frac{1}{k_{p,3}^2}\left\{ik_0\epsilon_3k_{p,3}\left[\sum_n{\tilde{C}_n^1H_{m-n}^{(1,2)}(k_{p,3}h)J'_m(k_{p,3}r_2)}+\tilde{C}_m^2H_m^{'(1,2)}(k_{p,3}r_2)\right]\right.-\\
-\left.\frac{k_zm}{r_2}\left[\sum_n{D_n^1}H_{m-n}^{(1,2)}(k_{p,3}h)J_m(k_{p,3}r_2)+D_m^2H_m^{(1,2)}(k_{p,3}r_2)\right]\right\}.
\end{split}
\end{equation}
Here $\tilde{A}_n^{1,2}=\sqrt{\epsilon_0/\mu_0}A_n^{1,2}$, $\tilde{C}_n^{1,2}=\sqrt{\epsilon_0/\mu_0}C_n^{1,2}$.
\end{widetext}

\bibliography{apssamp}

\end{document}